\documentclass[twocolumn, twocolappendix]{aastex631} 
\usepackage{bm}
\usepackage{amsmath}
\usepackage{upgreek}
\usepackage{acronym}
\let\tablenum\relax	
\usepackage[print-unity-mantissa=false]{siunitx}
\usepackage{url}
\usepackage[normalem]{ulem}
\usepackage{appendix}
\usepackage{rotating}

\newcommand{\msun}{\mathrm{M}_\odot}
\newcommand{\zsun}{\mathrm{Z}_\odot}

\graphicspath{{./}{./}}

\newcommand{\Zsun}{\textnormal{Z}_\odot}

\acrodef{ms}[MS]{main sequence}
\acrodef{zams}[ZAMS]{zero-age main sequence}
\acrodef{tams}[TAMS]{terminal-age main sequence}

\begin{document}

\title{Evolution of the Convective Core Mass during the Main Sequence}

\correspondingauthor{Minori Shikauchi}
\email{minori.shikauchi1020@gmail.com}

\author[0000-0002-3561-8658]{Minori Shikauchi}
\affiliation{Japan Synchrotron Radiation Research Institute, 1-1-1 Kouto, Sayo, Hyogo, 679-5148, Japan}

\author[0000-0000-0000-0000]{Ryosuke Hirai}
\affiliation{Astrophysical Big Bang Laboratory (ABBL), Cluster for Pioneering Research, RIKEN, Wako, Saitama 351-0198, Japan}
\affiliation{School of Physics and Astronomy, Monash University, Clayton, VIC3800, Australia}
\affiliation{OzGrav: The ARC Centre of Excellence for Gravitational Wave Discovery, Clayton, VIC3800, Australia}

\author[0000-0000-0000-0000]{Ilya Mandel}
\affiliation{School of Physics and Astronomy, Monash University, Clayton, VIC3800, Australia}
\affiliation{OzGrav: The ARC Centre of Excellence for Gravitational Wave Discovery, Clayton, VIC3800, Australia}



\begin{abstract}
We construct a semi-analytical model that describes the convective core mass evolution of massive stars experiencing mass loss during the main-sequence stage.
We first conduct a suite of 1D stellar evolution calculations to build insight into how convective core masses behave under idealized mass loss. Based on these simulations, we find several universal relations between global properties of the star that hold regardless of the mass loss history. By combining these relations, we construct a semi-analytic framework that can predict the convective core mass evolution for arbitrary mass loss histories and hence the helium core mass at the end of the main sequence.
Our formulae improve upon existing methods for predicting the core mass in rapid population synthesis codes. 

\end{abstract}

\keywords{stellar evolution (1599) --- binary stars (154)}


\section{Introduction} \label{sec:intro}
Massive stars burn hydrogen via the CNO cycle during their main-sequence phase, and due to the strong temperature dependence of the nuclear reactions, their cores are convectively unstable. The convective motions strongly mix the processed elements over the entire core and the size of this region generally shrinks throughout the main-sequence phase, as the mean molecular weight increases over time and the opacity decreases. The main-sequence phase ends when the core runs out of hydrogen fuel, i.e., the whole convective core consists of helium. The subsequent evolution of the star is mostly dictated by this helium core mass: the final fate (core-collapse supernova vs collapse to black hole vs pair-instability supernova, etc.), and the mass and type of remnant it leaves (white dwarf, neutron star, black hole, no remnant), etc. 

The size of the convective core and how it evolves mostly depends on the physics of convection (e.g. convective boundary mixing). However, the core evolution can also be strongly altered when the star loses mass. Massive stars on the main sequence can lose mass for a number of reasons such as stellar winds, eruptions and binary interactions. In particular, binary interactions can strip mass off stars at different rates and at different times depending on the binary properties, so there can be infinite variations to the mass-loss history. Therefore, it is entirely possible that two stars with identical masses at the point of core hydrogen depletion have different helium core masses depending on their mass-loss histories during the main-sequence phase. 

In this paper, we investigate how convective cores evolve during the main-sequence phase. We mainly explore the massive star range where black holes are formed at the end of their lives, as the black hole masses are expected to be directly tied to the helium core masses. We particularly focus on how the convective core mass responds to different degrees of mass loss occurring at various epochs. By carrying out a suite of 1D stellar evolution simulations with idealized mass loss, we find some useful relations that apply regardless of the detailed mass-loss histories. Based on these findings, we construct a semi-analytical model that can accurately predict the evolution of convective cores for arbitrary stars with arbitrary mass loss. 

Such simplified prescriptions for predicting convective core mass evolution could be useful for a range of applications. For example, the maximum achievable mass of isolated stellar mass black holes strongly depends on the mass loss in winds \citep[e.g.][]{Belczynski2010,Romagnolo:2024,Vink2024}. Our proposed method will allow us to test various mass-loss prescriptions with very low computational cost. Another example arises in binary evolution. Previous studies showed that the method used in many rapid binary population synthesis codes to predict the core mass at \ac{tams} \citep{Hurleyetal2000, Hurleyetal2002} severely underestimates the core mass, especially when the stellar mass at \ac{tams} is largely different to its \ac{zams} value due to binary interactions \citep{Neijsseletal2021, Romero-Shawetal2023,Bavera:2023}. This can have significant impacts on predictions for X-ray binaries and gravitational-wave sources. While some procedures have been proposed to remedy this issue \citep{Neijsseletal2021, Romero-Shawetal2023}, our new model grounded in detailed 1D stellar evolution calculations provides a much more accurate prediction.

The structure of this paper is as follows.
In Section~\ref{sec:method}, we outline the numerical method for our stellar evolution calculations.
Section~\ref{sec:result} introduces some universal relationships obtained from the simulations and a semi-analytical framework to describe the convective core mass evolution under arbitrary mass loss histories.
We discuss the robustness of our framework and possible applications in Section~\ref{sec:discussion}.
We summarize our conclusions in Section~\ref{sec:conclusion}.

\section{Method} \label{sec:method}

In this section, we outline our numerical method for simulating the main-sequence phase of stellar evolution under idealized mass-loss histories. We do not specify the origin of the mass loss, whether it is due to stellar winds, eruptions or binary interactions. This simplified approach allows us to grasp the fundamental features of the convective core response to mass loss.

We perform a suite of 1D stellar evolution calculations of the main-sequence phase using the public code MESA \citep{Paxtonetal2011, Paxtonetal2013, Paxtonetal2015, Paxtonetal2018, Paxtonetal2019, Jermynetal2023} v.23.05.1. Most evolution parameters are set to default values. The Ledoux criterion is used to identify convective stability and mixing length theory is used to treat convective energy transport. For our fiducial model, we use a mixing length parameter $\alpha_\mathrm{mlt}=2$. Step overshooting is applied above the convective core boundary with an overshoot parameter $\alpha_\mathrm{ov}=0.2$. The inlists and subroutines used in our simulations can be found in \url{https://zenodo.org/doi/10.5281/zenodo.12662373}. 

In order to isolate the effect of mass loss on the convective core evolution, we carry out simulations with and without mass loss for comparison. For the non-mass-loss models, we use 11 values of \ac{zams} masses in the range $M_\mathrm{zams}\in [15,100]~\msun$ (see Table \ref{tab:params}). We also run simulations with mass loss for the models with \ac{zams} masses $M_\mathrm{zams}\in [15,60]~\msun$. In mass-losing models, a constant mass loss rate is applied starting at the time when the central helium mass fraction $Y_\mathrm{c}$ reaches a specified value $Y_\mathrm{c, ml}=Y_{\mathrm{c,zams}}+\tau_\mathrm{ml}X_{\mathrm{c,zams}}$. Here, $X_{\mathrm{c,zams}}$ and $Y_{\mathrm{c,zams}}$ are the central hydrogen and helium fractions at \ac{zams} respectively, and $\tau_\mathrm{ml} \in (0,1)$ is a dimensionless parameter that expresses the fractional evolutionary age of the star when we start to apply mass loss.
Three different mass loss rates are applied, $\dot{M}=-5 \times 10^{-6},\ -1 \times 10^{-5},\  $\qty{-1.5e-5}{\msun. yr^{-1}}. These correspond to a mass-loss timescale of several \qty{}{\mega yr}, which is roughly of the order of the nuclear timescale of stars in the mass range we explore. We terminate the mass loss once the total stellar mass is reduced to the mass of the ``mixing core'' (similar to the convective core; see below for definition of ``mixing core'') at \ac{zams}. The stellar radii start to rapidly drop at this point due to the change in surface composition, so this very roughly corresponds to when binary mass transfer processes would cease and rapid wind mass loss, particularly through luminous blue variable winds, would reduce. All simulations are then continued until \ac{tams}.
We also explore three different values of metallicity, $Z=\Zsun,\  \Zsun/3,\ \Zsun/10$, where the solar value is defined as $\Zsun=0.02$. All parameter variations considered in this work are summarized in Table~\ref{tab:params}.

In our simulations, we define a ``mixing core'' that includes both the convective core and its adjacent overshoot region. The mixing core characterizes the central region of the star where it is fully chemically mixed, which directly determines the helium core mass at the end of the main-sequence phase. Typically, this is a few to 10 per cent larger than the convective core mass. The main focus of these simulations is to track the evolution of the mixing core mass $m_\mathrm{mix}$ over time with and without mass loss, along with the stellar luminosity and the central helium mass fraction.

\begin{table*}[ht!]
	\centering
\begin{tabular}{cccccc} \hline
Symbol & Definition & Values \\ \hline
$M_\mathrm{zams}$ [$\msun$] & initial total mass & $15, 20, 25, 30, 40, 50, 60, 70, 80, 90, 100$ \\
$\dot{M}$ [\qty{}{\msun. yr^{-1}}]$^{*}$ & mass loss rate & $-5 \times 10^{-6}, -1 \times 10^{-5}, -1.5 \times 10^{-5}$ \\
$\tau_\mathrm{ml}^{*}$ & mass loss onset time & $0.2, 0.4, 0.6$ \\
$Z$ [$\zsun$] & initial metallicity  & $0.1, 1/3, 1.0$ \\ \hline
\end{tabular}
	\caption{Summary of parameters used for our MESA simulations.  $^*$Mass loss is applied only to stars with masses $M_\mathrm{zams}\leq60~\msun$.
 }
 \label{tab:params}
\end{table*}


\section{Results} \label{sec:result}

All models presented in this section are based on our fiducial set of evolution parameters and metallicity ($\alpha_\mathrm{mlt}=2, \alpha_\mathrm{ov}=0.2, Z=\Zsun$) unless stated otherwise, but the same qualitative relations hold for other combinations of these parameters.

\subsection{General behaviour of core mass evolution}
\begin{figure}
\includegraphics[width=\linewidth]{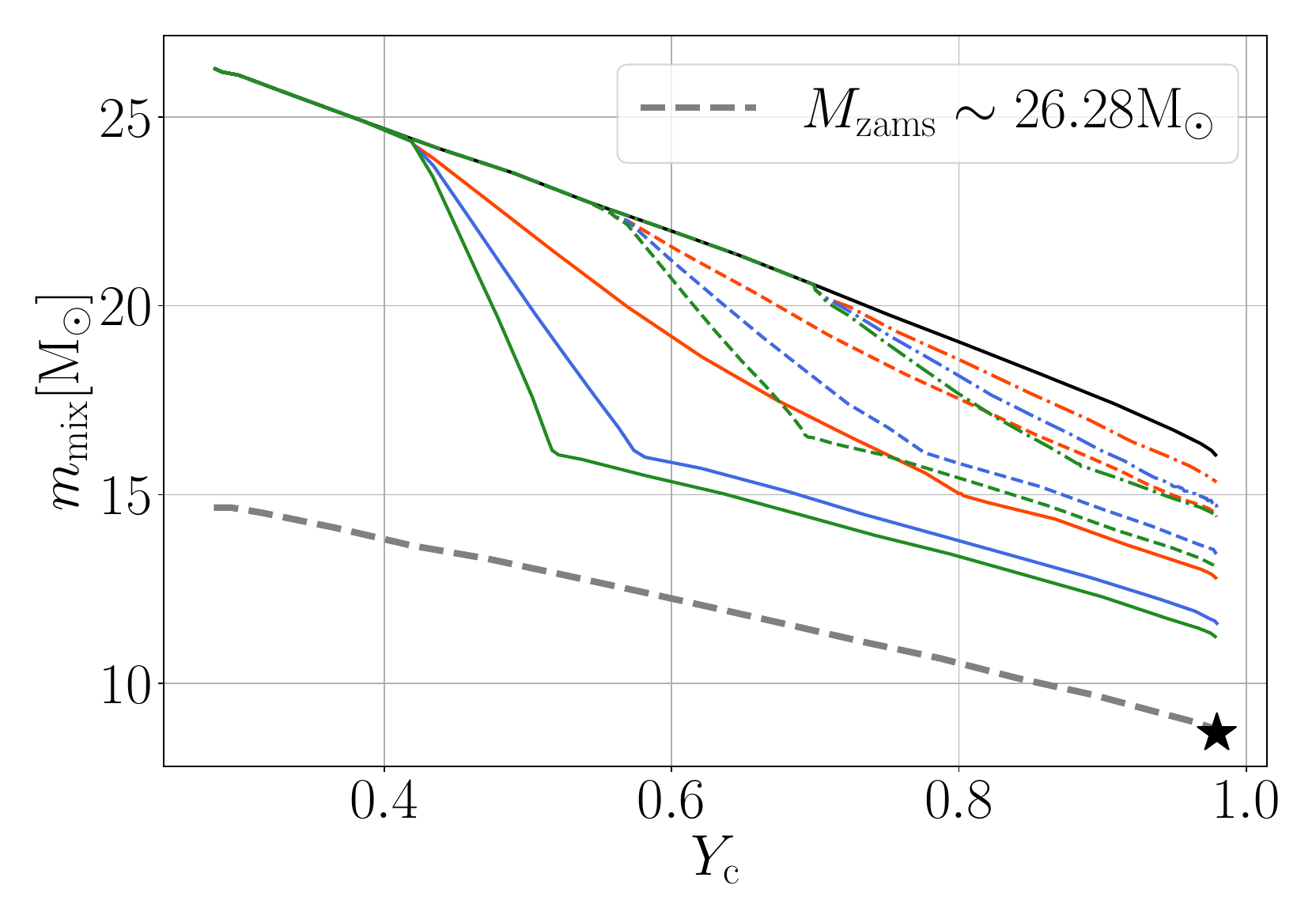}
\caption{Core mass evolution as a function of the central helium fraction. Colored lines correspond to the evolution of a star with an initial total mass of \qty{40}{\msun} but different mass loss rates and timings. Each color indicates different mass loss rates: no mass loss (black), \qty{-5e-6}{\msun. yr^{-1}}(orange), \qty{-1e-5}{\msun. yr^{-1}} (blue), and \qty{-1.5e-5}{\msun. yr^{-1}} (green), while the line styles show different mass loss onset times: $\tau_\mathrm{ml}=0.2$ (solid), $0.4$ (dashed) and $0.6$ (dash-dotted). For all the models, mass loss ceases when the total mass reaches the initial mixing core mass, $\sim$\qty{26.28}{\msun}. For comparison, the grey dashed curve shows a model evolved from $M_\mathrm{zams}=$\qty{26.28}{\msun}, which is the final total mass for the mass-losing models. Note that for some mass-losing models the main-sequence phase terminates before reaching the final intended mass.}
\label{fig:comp_bse}
\end{figure}

Figure~\ref{fig:comp_bse} depicts the mixing core mass evolution obtained from our MESA simulations. We use $Y_\mathrm{c}$ as a measure of age, as it monotonically increases from the primordial value to $Y_\mathrm{c}=1-Z$ over the duration of the main-sequence phase as long as there is no mass gain.
Each colored curve shows the evolution for a fixed initial total mass $M_\mathrm{zams}=40~\msun$, with a variety of mass loss histories ($\tau_\mathrm{ml}, \dot{M}$). In general, the mixing core mass declines steadily throughout the evolution. There is an abrupt change in the slope of the decline as mass loss is switched on, with steeper slopes for higher mass loss rates. The slope turns shallow again once mass loss is switched off.

An important point to be noted is that the mixing core mass at \ac{tams} is different for each model. All cases end up with core masses that are lower than that of the model without mass loss (solid black curve) and larger than the model evolved without mass loss from $M_\mathrm{zams}\sim26.28~\msun$ (grey dashed curve), which is the final total mass for the mass-losing models. This highlights that the full mass loss history is relevant for determining the helium core mass at \ac{tams}. Moveover, this clearly indicates that the common approach taken in rapid population synthesis codes based on \cite{Hurleyetal2000,Hurleyetal2002} underestimates the core mass, as it predicts that the helium core mass at \ac{tams} depends only on the total stellar mass at \ac{tams}.

\subsection{Universal relations in convective core evolution} \label{subsec:univ_rel}
In this section, we present some universal relations we find within our stellar evolution models. These relations prove useful in constructing an analytical framework for modelling the evolution of the mixing core. 

\begin{figure}
\includegraphics[width=\linewidth]{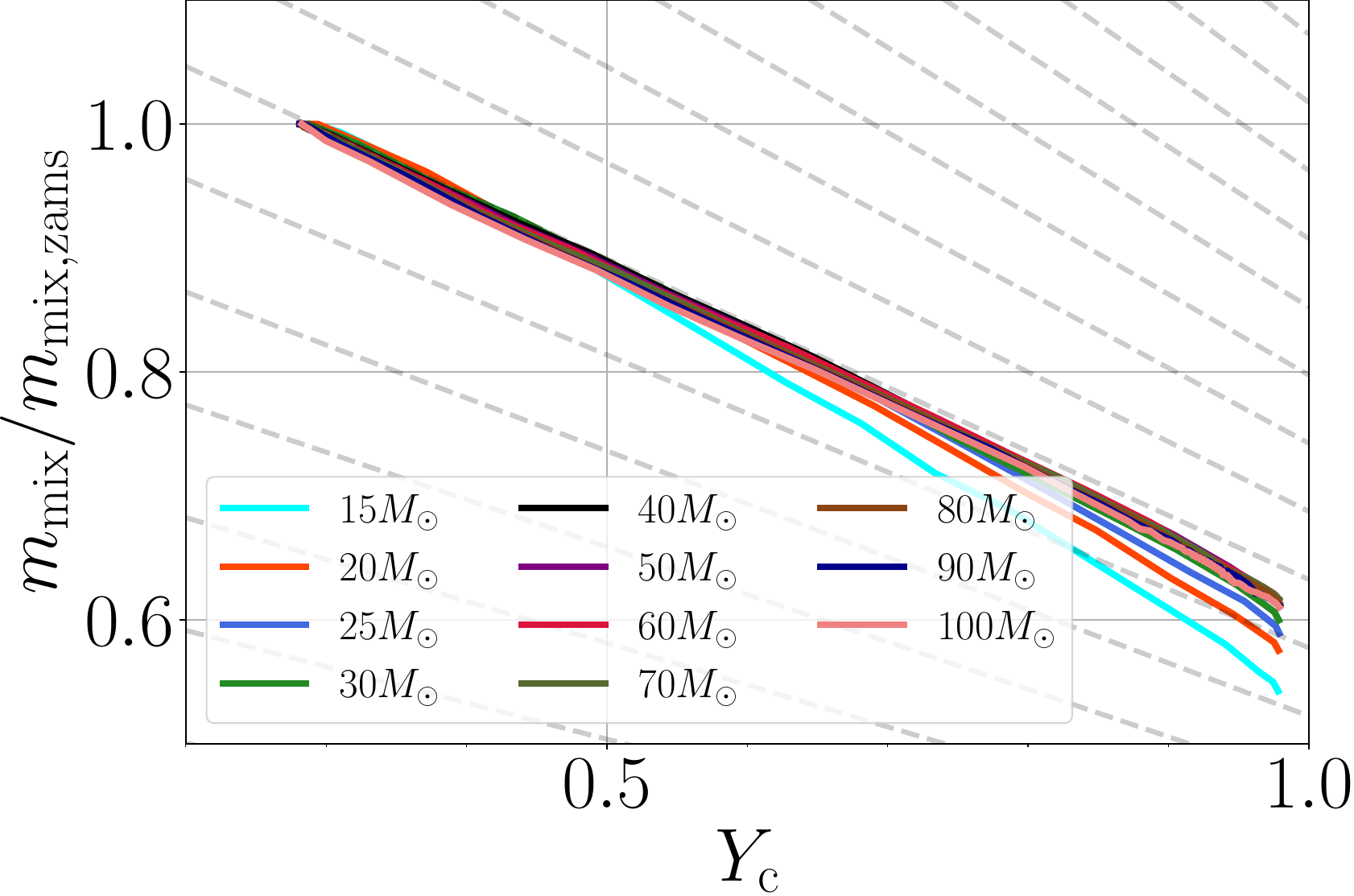}
\caption{Mixing core mass evolution as a function of the central helium fraction for the non-mass-losing models. The mixing core mass is normalized by its initial value, $m_\mathrm{mix,zams}$.  Each colored line corresponds to a different value of initial mass, as stated in the legend. Grey lines correspond to mixing core mass evolution with constant $\alpha$ (see text for definition).}
\label{fig:all_mmix}
\end{figure}

\subsubsection{Natural decrease rate of core mass with central helium fraction}
The first relation we focus on is the rate of the decrease in the mixing core mass in the absence of mass loss.
Figure~\ref{fig:all_mmix} plots the mixing core mass $m_\mathrm{mix}$ against central helium fraction $Y_\mathrm{c}$ in stars with various initial total masses without mass loss. Mixing core masses are normalized by their values at \ac{zams}.  It is well known that for stars that have convective cores, the size of the convective region decreases over time due to the decrease in electron scattering opacity as hydrogen is converted into helium. Here, for stars with an initial total mass exceeding \qty{30}{\msun}, the core masses exhibit a roughly universal monotonic linear decline, i.e.,
\begin{equation}
    \frac{1}{m_\mathrm{norm}} \frac{\mathrm{d}m_\mathrm{mix}}{\mathrm{d}Y_\mathrm{c}}\sim\mathrm{const.},
    \label{mmix_const}
\end{equation}
where $m_\mathrm{norm}$ is a normalization mass. In this case, it is the mixing core mass at \ac{zams}, i.e.~$m_\mathrm{norm}=m_\mathrm{mix,zams}$. This means that the ratio of the initial to final mixing core mass is roughly universal ($m_\mathrm{mix,tams}/m_\mathrm{mix,zams}\sim0.6$) between stars of different masses. There are some small deviations from the trend only in the lower mass models, where the slope $\mathrm{d}m_\mathrm{mix}/\mathrm{d}Y_\mathrm{c}/ m_\mathrm{norm}$ becomes steeper at later times in the evolution.

\begin{figure}[ht!]
\includegraphics[width=\linewidth]{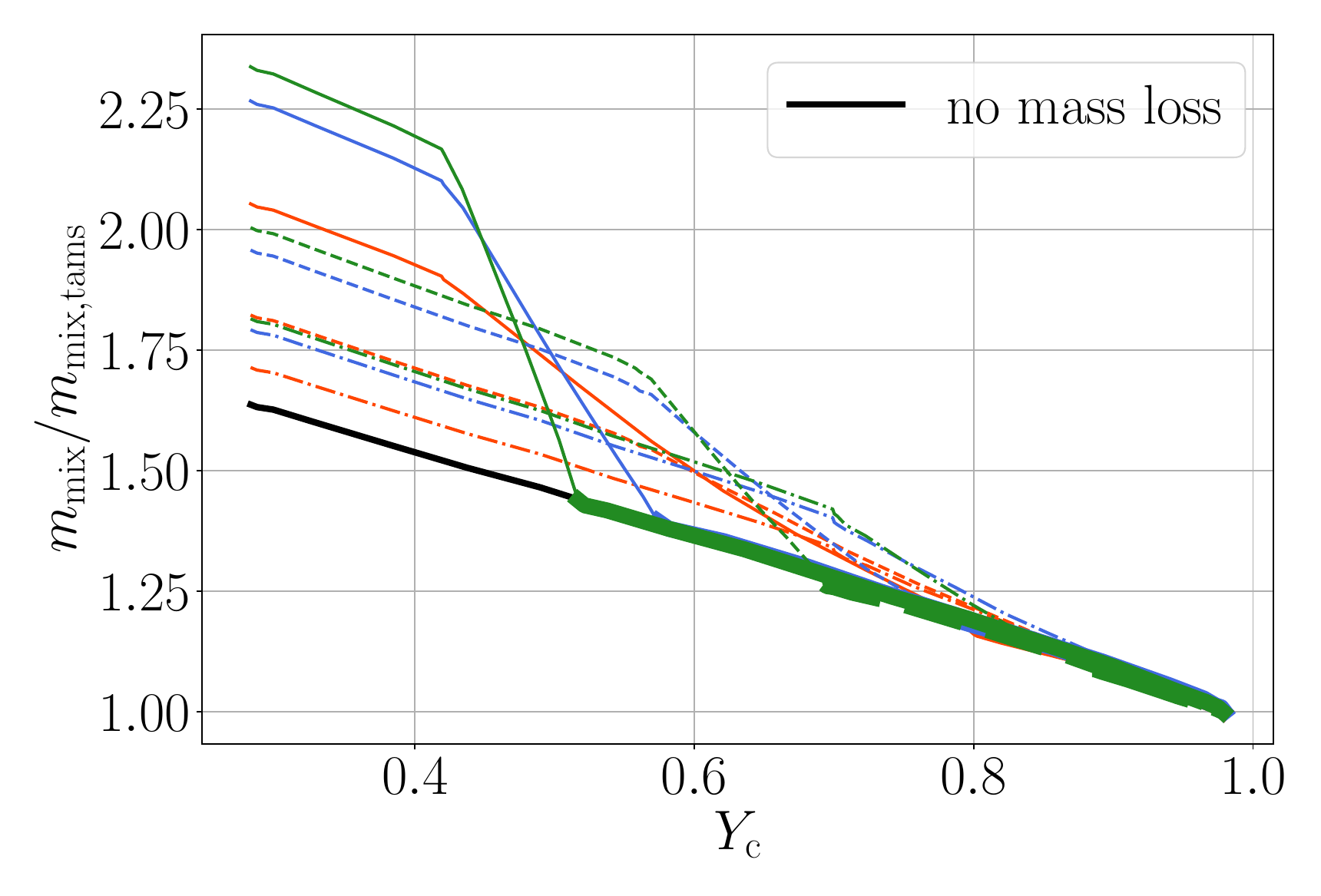}
\caption{Mixing core mass evolution for a \qty{40}{\msun} star. The black line corresponds to the case without mass loss.   The core mass is normalized by the value at \ac{tams} ($Y_\mathrm{c} \sim 0.98$). Line colours and styles are defined in the same way as in Figure~\ref{fig:comp_bse}.
\label{fig:align}}
\end{figure}

We further find that this trend is preserved even after mass loss.
Figure~\ref{fig:align} shows core mass evolution with core mass normalized by the value at \ac{tams}.
If we focus on the thick portions of the curves, which correspond to the evolution after mass loss is switched off, we see that the decline is again linear and the slopes closely agree between models with different total masses and mass loss rates. In these regions, Equation~(\ref{mmix_const}) holds again by taking $m_\mathrm{norm}=m_\mathrm{mix,tams}$. We already show above that the ratio $m_\mathrm{mix,tams}/m_\mathrm{mix,zams}$ is roughly constant in the absence of mass loss. Therefore, the relation is equivalent to saying that we are replacing $m_\mathrm{norm}=m_\mathrm{mix,zams}$ with an effective initial mass $m_\mathrm{mix,zams,eff}\sim m_\mathrm{mix,tams}/0.6$. 

These relations indicate that regardless of the \ac{zams} mass and mass loss history, the decline rate of the mixing core mass in the absence of ongoing mass loss is solely determined by the current $m_\mathrm{mix}$ and $Y_\mathrm{c}$. Based on these findings, we define the following value:
\begin{equation}
    \alpha \equiv -\left(1 - \frac{\mathrm{d} \ln m_\mathrm{mix}}{\mathrm{d} \ln Y_\mathrm{c}}\right)^{-1}\frac{\mathrm{d} \ln m_\mathrm{mix}}{\mathrm{d} Y_\mathrm{c}}.\label{eq:alpha_def}
\end{equation}
The right hand side is equivalent to Equation~(\ref{mmix_const}) by substituting
\begin{equation}
 m_\mathrm{norm}=m_\mathrm{mix}-\frac{\mathrm{d}m_\mathrm{mix}}{\mathrm{d}Y_\mathrm{c}}\times Y_\mathrm{c}.
\end{equation}
Here, $m_\mathrm{norm}$ corresponds to the mixing core mass at $Y_\mathrm{c}=0$, if we extrapolate back with the universal slope from the current mixing core mass. In our fiducial set of models, $\alpha\sim 0.45$ almost universally except for small deviations at lower masses and later in the evolution. Using this value, we can express the ``natural'' decline rate of the mixing core mass as
\begin{equation}
 \left.\frac{\mathrm{d}\ln m_\mathrm{mix}}{\mathrm{d}Y_\mathrm{c}}\right|_\mathrm{nat}=-\frac{\alpha}{1 - \alpha Y_\mathrm{c}},\label{eq:natural_decline_rate}
\end{equation}
which holds for any massive star with no ongoing mass loss.

Although we have built this expression based on the observation that $\alpha\sim\mathrm{const.}$ for most of the parameter region, we also made a fit for the value of $\alpha$ as a function of $m_\mathrm{mix}$ to account for the small deviations. See Appendix \ref{app:formula_fid} for details of our fitting function. In the following we assume that the natural decline rate can be expressed by Eq.~(\ref{eq:natural_decline_rate}) with the fitted form of $\alpha=\alpha(m_\mathrm{mix})$.

\subsubsection{Relation between luminosity, mixing core mass and central helium fraction}\label{sec:luminosity}
\begin{figure}[ht!]
\includegraphics[width=\linewidth]{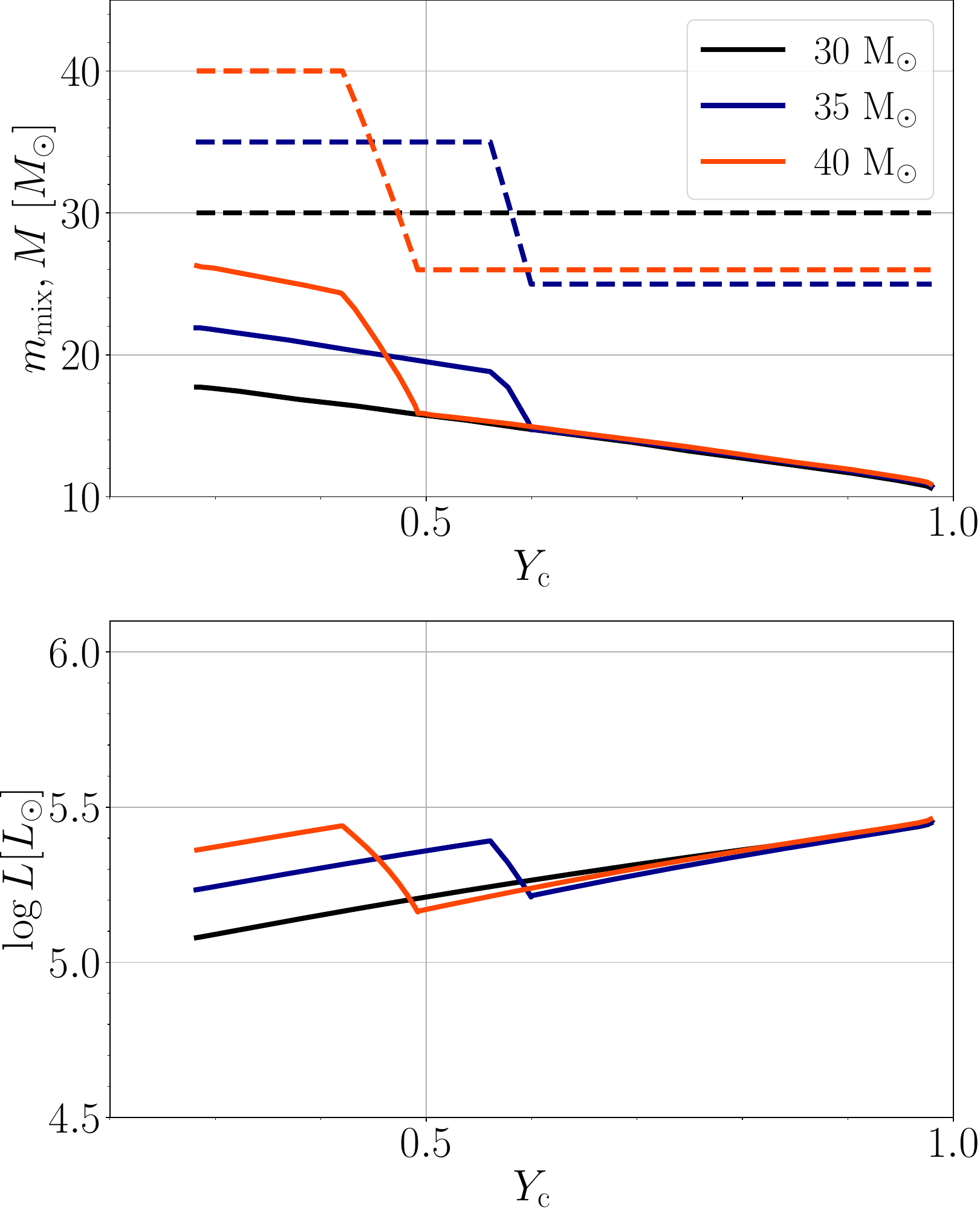}
\caption{Evolution of mixing core mass and luminosity as a function of central helium fraction with different choices of mass loss history. In the upper panel, the mixing core and total mass evolution are shown with solid and dashed curves, respectively. Colors denote the \ac{zams} mass, as shown in the legend.  The mass loss rates and durations are tuned so that the mixing core masses evolve in the same way after mass loss ceases. }
\label{fig:luminosity_alignment}
\end{figure}

\begin{figure}[ht!]
\includegraphics[width=\linewidth]{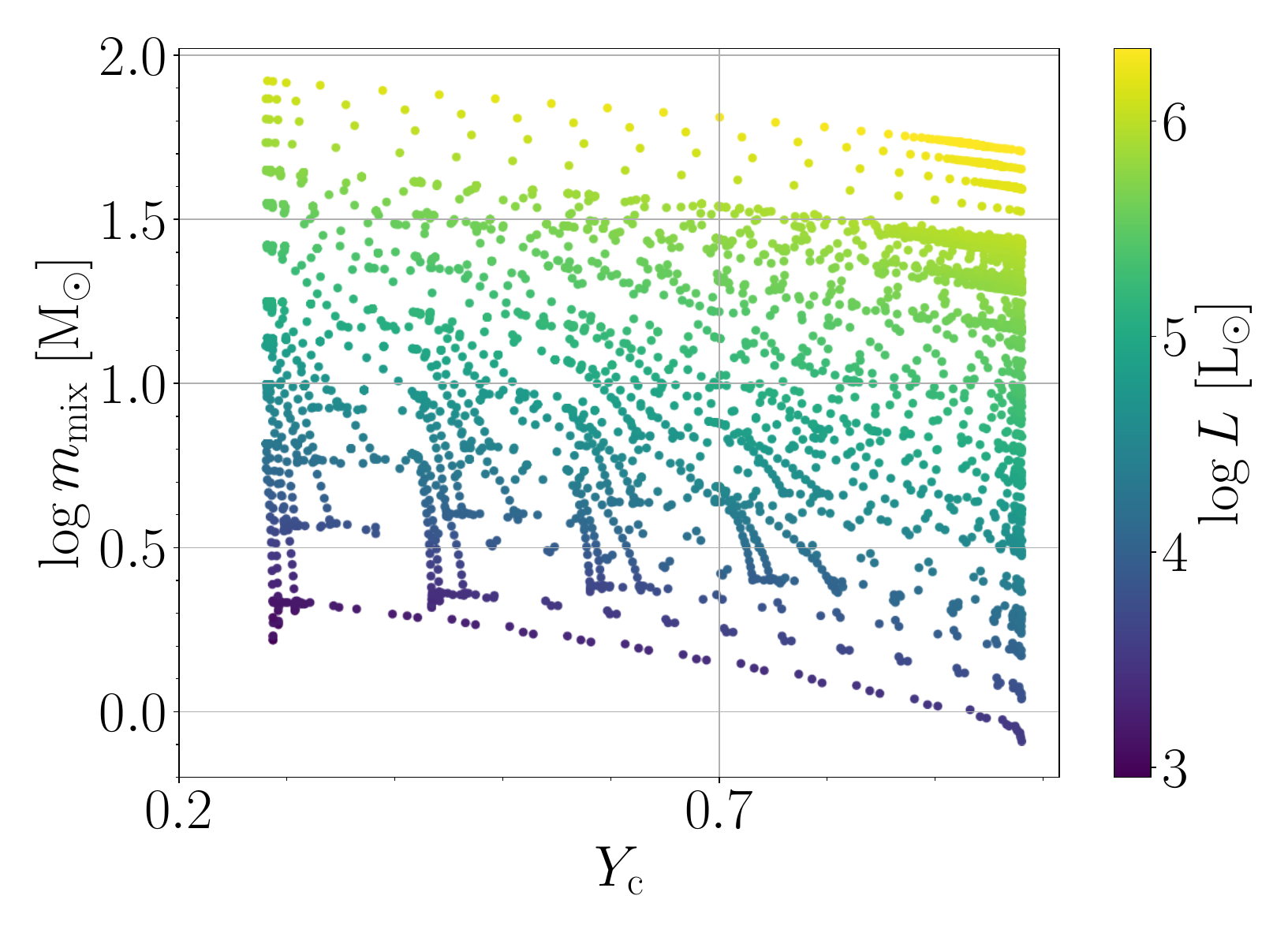}
\caption{Luminosity of the star as a function of the mixing core mass and central helium fraction. All the data obtained in the simulations are shown, with different ZAMS masses, mass loss timings and mass loss rates. Each point corresponds to a snapshot in the MESA models. The points are colored by the value of luminosity.  
\label{fig:L_mmix}}
\end{figure}

Another important relation we find is a strong correlation between the mixing core mass $m_\mathrm{mix}$, central helium fraction $Y_\mathrm{c}$ and luminosity of the star $L$. In Figure~\ref{fig:luminosity_alignment} we compare three different models with different \ac{zams} masses and mass loss histories. The mass loss onset/switch-off ages were tuned so that the mixing core masses agree with each other once mass loss is switched off (upper panel). Despite having different total masses and chemical structures, the luminosities depend almost exclusively on $m_\mathrm{mix}$ and $Y_\mathrm{c}$
(lower panel).

Figure~\ref{fig:L_mmix} shows the luminosity as a function of $m_\mathrm{mix}$ and $Y_\mathrm{c}$ for all the models we computed. Each point corresponds to a single timestep in our MESA models. We see that the distribution of $L$ is smooth in this space despite having models with different \ac{zams} masses and mass loss histories plotted on top of each other. This strengthens our claim above that $L$ is only a function of the current $m_\mathrm{mix}$ and $Y_\mathrm{c}$, and all other factors (total mass, chemical structure, etc.) only play minor roles.

Physically, this can be interpreted as follows.
The convective core mass is the mass coordinate of the boundary between where the convective stability criterion is satisfied and not. Convective stability is determined by the Ledoux criterion in our models, but given that the convective boundary has a flat chemical profile due to overshooting, this reduces to the Schwarzschild criterion. The quantities used in the Schwarzschild criterion are the local luminosity, mass coordinate, opacity, pressure and temperature. Among these quantities, the opacity closely correlates with $Y_\mathrm{c}$ as in the central regions of a main sequence star, electron scattering is the dominant opacity source and depends only on the hydrogen fraction.
The other four quantities -- luminosity, mass, pressure and temperature -- are not completely independent from each other. Main sequence stars can be assumed to be in complete (hydrostatic+thermal) equilibrium, so the two structure equations reduce the number of independent variables from four to two. This explains the strong correlation between the three variables $m_\mathrm{mix}, L$ and $Y_\mathrm{c}$, which we utilize in the later sections. In Appendix~\ref{app:formula_fid}, we present fitting formulae for $L=L(m_\mathrm{mix},Y_\mathrm{c})$. 

\subsubsection{Core mass response to mass loss} \label{subsec:w_mt}
\begin{figure}[ht]
\includegraphics[width=\linewidth]{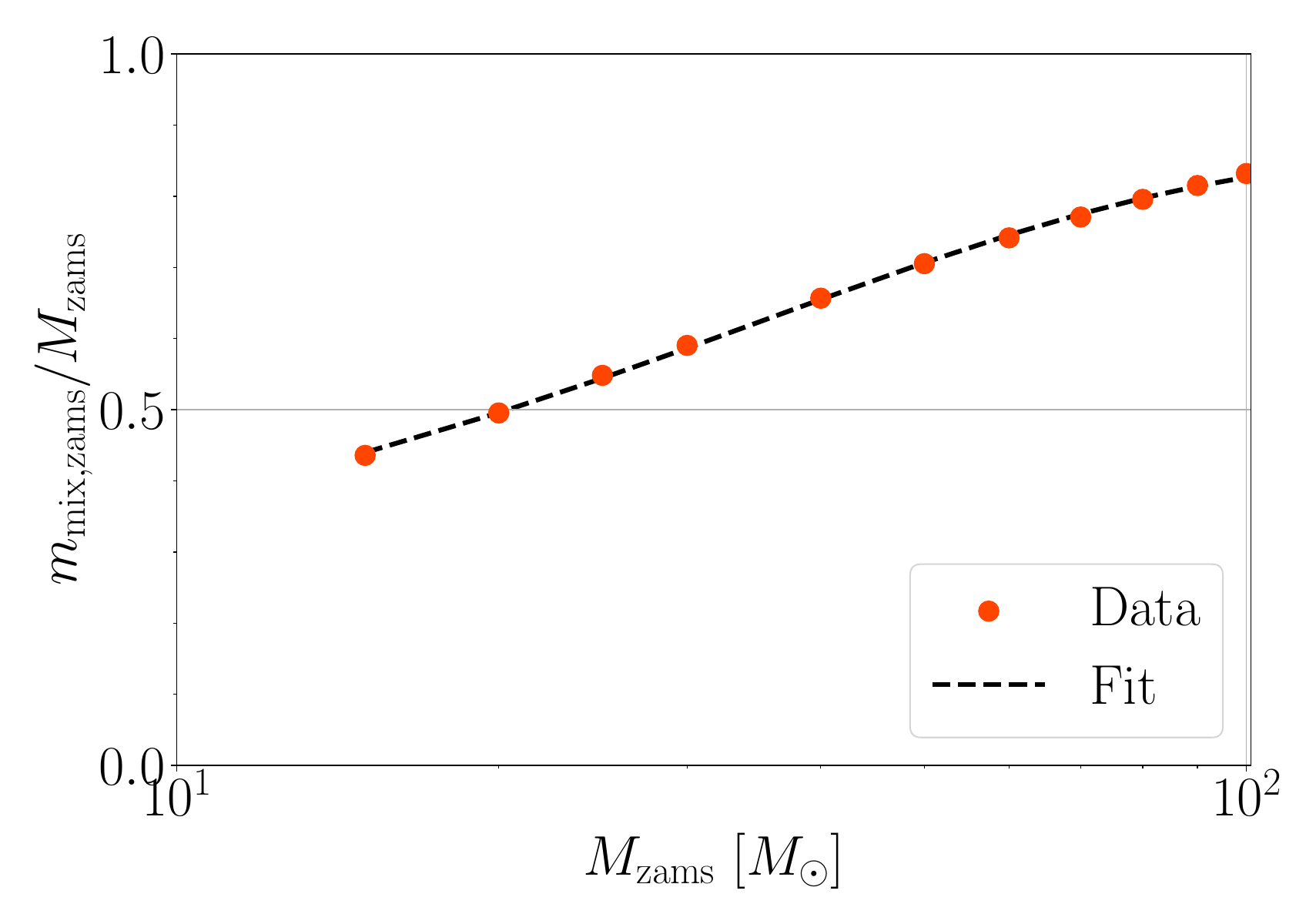}
\caption{Ratio of $m_\mathrm{mix,zams}$ to $M_\mathrm{zams}$ from the MESA models (orange dots). The dashed black curve shows the fitting formula given by Eq.~(\ref{fmixfit}) in Appendix~\ref{app:formula_fid}.}
\label{fig:beta}
\end{figure}

In the previous sections we have found a universal expression for the mixing core mass decline rate in the absence of mass loss. Here, we investigate the effect of mass loss on the mixing core evolution. To isolate the effect of mass loss, we assume that the mixing core mass decline rate is determined as a linear sum of the natural decline and a term driven by the mass loss. Using $Y_\mathrm{c}$ as a measure of age, this can be expressed as
\begin{equation}
 \frac{\mathrm{d} m_\mathrm{mix}}{\mathrm{d} Y_\mathrm{c}} = \left . \frac{\mathrm{d} m_\mathrm{mix}}{\mathrm{d} Y_\mathrm{c}} \right|_\mathrm{nat}+\left . \frac{\mathrm{d} m_\mathrm{mix}}{\mathrm{d} Y_\mathrm{c}} \right|_\mathrm{ml}.
 \label{eq:dmdY_sum}
\end{equation}
The first term on the right hand side represents the natural decrease in the core mass without the mass loss, while the second term accounts for the impact of mass loss.
There are several conditions that the second term on the right hand side should satisfy. First, it should be proportional to the mass loss rate since it vanishes in the absence of mass loss:
\begin{equation}
 \left . \frac{\mathrm{d} m_\mathrm{mix}}{\mathrm{d} Y_\mathrm{c}} \right|_\mathrm{ml}\propto \dot{M},
\end{equation}
where $M$ is the total mass of a star.
Secondly, at \ac{zams} and in the limit of a high mass-loss rate, the mass of the mixing core after mass loss should follow that of a \ac{zams} star of the given total mass, \textit{i.e.}
\begin{equation}
 \left. \frac{\mathrm{d}m_\mathrm{mix}}{\mathrm{d}Y_\mathrm{c}}\right|_\mathrm{ml}(Y_\mathrm{c}=Y_\mathrm{c,zams})\propto \frac{\mathrm{d}m_\mathrm{mix,zams}}{\mathrm{d}M_\mathrm{zams}}.
\end{equation}
In Figure~\ref{fig:beta}, we show the fraction of the mass contained in the mixing core ($f_\mathrm{mix}\equiv m_\mathrm{mix,zams}/M_\mathrm{zams}$) for \ac{zams} stars with  different masses. It shows that the fractional mass of the convective core is higher for higher mass stars. If the mass loss rate from a \ac{zams} star is high enough that the change in $Y_\mathrm{c}$ is negligible over the duration of mass loss, the star essentially stays as a \ac{zams} star so the mixing core mass should follow this curve in response to mass loss.

Given these constraints, we propose that the contribution of mass loss to the mixing core mass evolution can be described by the following expression
\begin{equation}
 \left. \frac{\mathrm{d} \ln m_\mathrm{mix}}{\mathrm{d} Y_\mathrm{c}} \right|_\mathrm{ml} = \beta(M) \delta \frac{\mathrm{d} \ln M}{\mathrm{d} Y_\mathrm{c}},\label{eq:dmdy_model}
\end{equation}
where $\beta(M_\mathrm{zams})$ is 
\begin{equation}
 \beta(M_\mathrm{zams}) \equiv \frac{\mathrm{d} \ln m_\mathrm{mix, zams}}{\mathrm{d} \ln M_\mathrm{zams}},\label{eq:beta}
\end{equation}
which can be derived from the relationship between $m_\mathrm{mix,zams}$ and $M_\mathrm{zams}$ shown in Figure~\ref{fig:beta}. We present a fitting formula for $f_\mathrm{mix}(M_\mathrm{zams})$ in Appendix~\ref{app:formula_fid} (black dashed curve in Figure~\ref{fig:beta}), which can be used to compute $\beta$.
The factor $\delta$ contains all the other effects that we have not taken into account with the other factors. It can be interpreted as quantifying the \textit{inertia} of the mixing core to mass loss, where lower values of $\delta$ mean that the core responds less to the decrease in total mass.

To facilitate this analysis, we introduce a new variable, $\hat{Y}_\mathrm{c}$, defined as
\begin{equation}
    \hat{Y}_\mathrm{c} \equiv \frac{Y_\mathrm{c} - Y_\mathrm{c,zams}}{1 - Y_\mathrm{c,zams} - Z}
    = \frac{Y_\mathrm{c} - Y_\mathrm{c,zams}}{X_\mathrm{c,zams}},
\end{equation}
which linearly maps the central helium fraction; $\hat{Y}_\mathrm{c}$ grows from 0 to 1 during the main sequence. The functional form of $\delta$ should be such that $\delta=1$ at $\hat{Y}_\mathrm{c}=0$ and $\delta$ declines as the central helium fraction builds up ($\hat{Y}_\mathrm{c}\rightarrow1$) and the mixing core mass becomes less sensitive to mass loss.

\begin{figure}
\includegraphics[width=\linewidth]{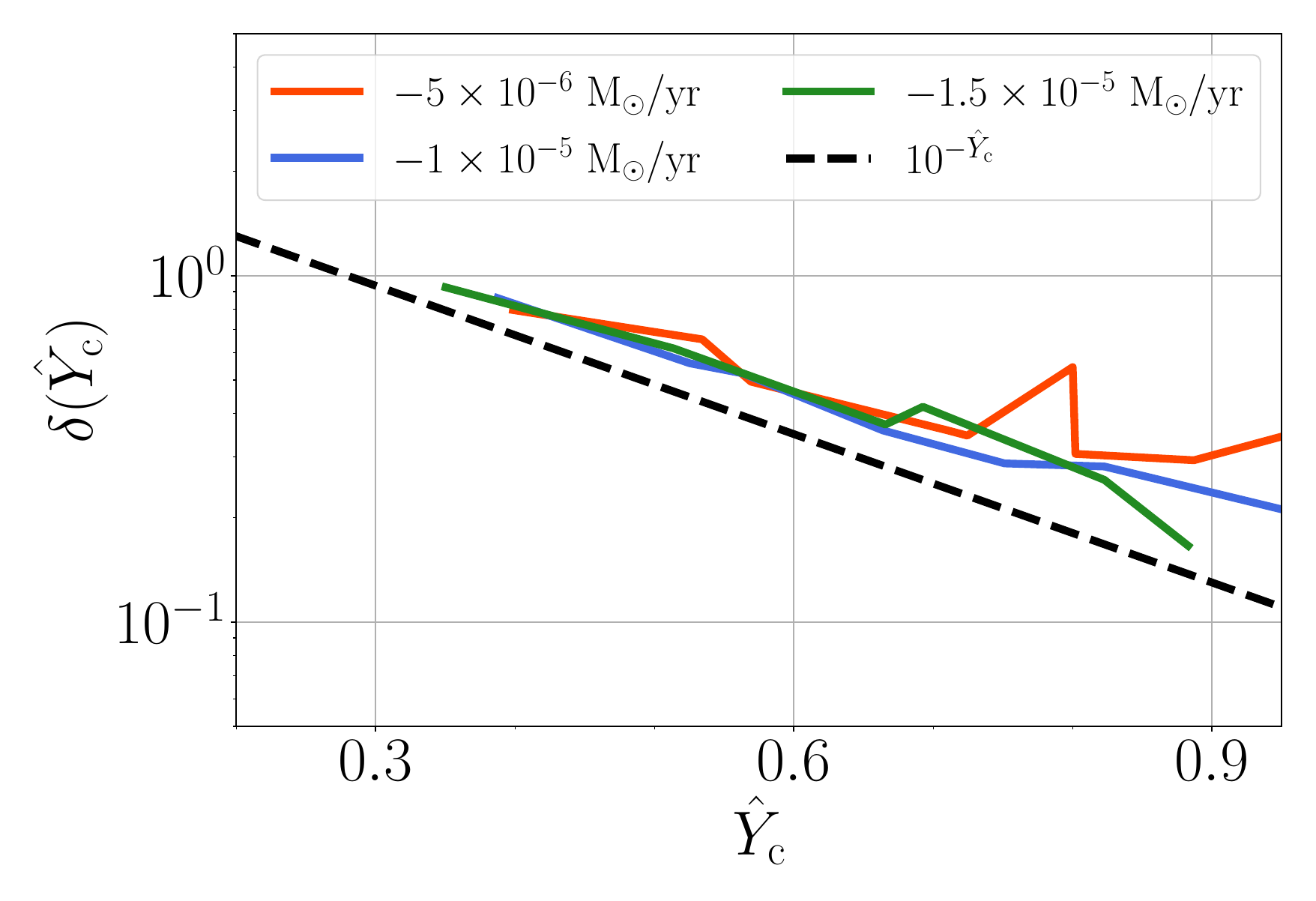}
\caption{The value of $\delta$ as a function of $\hat{Y}_\mathrm{c}$. Each colored solid line corresponds to a different value of the mass loss rate for a star with the same \ac{zams} mass $M_\mathrm{zams}=40~\msun$. As a reference, the dashed black line shows $\delta \propto 10^{-\hat{Y}_\mathrm{c}}$.}
\label{fig:delta}
\end{figure}

We already have a model for the natural decline rate (Eq.~(\ref{eq:natural_decline_rate})) and $\beta$ (Eq.~(\ref{eq:beta})), so given the numerical derivatives $d\ln m_\mathrm{mix}/dY_\mathrm{c}$ and $\mathrm{d}\ln M/\mathrm{d}Y_\mathrm{c}$, we can compute the value of $\delta$ in our MESA models by solving for $\delta$ from Eqs.~(\ref{eq:dmdY_sum}) and (\ref{eq:dmdy_model}):
\begin{equation}
    \delta = \beta^{-1}(M)\left(\frac{\mathrm{d}\ln M}{\mathrm{d}Y_\mathrm{c}}\right)^{-1}\left(\frac{\mathrm{d}\ln m_\mathrm{mix}}{\mathrm{d}Y_\mathrm{c}}+\frac{\alpha}{1-\alpha Y_\mathrm{c}}\right).\label{eq:delta}
\end{equation}
Some examples are plotted in Figure~\ref{fig:delta}. Here we show the numerically evaluated values of $\delta$ for models with the same initial total mass ($M_\mathrm{zams}=40~\msun$) and different mass loss rates. Note that $\delta$ is only defined during mass loss. Remarkably, the curves closely agree with each other despite having different mass loss rates, implying that $\delta$ is mostly a function of $\hat{Y}_\mathrm{c}$ only. Furthermore, the value of $\delta$ declines exponentially as $\delta\sim10^{-\hat{Y}_\mathrm{c}}$ with a nearly constant offset. We find similar behaviours for models with different initial stellar masses, although the value of the offset depends on the initial mass. Grounded by these findings, we assert that $\delta$ can be universally expressed as a function of $M_\mathrm{zams}$\footnote{It is likely that the $M_\mathrm{zams}$-dependence is only a correlation and there is a more direct cause that can be expressed with variables of the current state of the star. For the current purpose, it is sufficient to leave it as dependent on the initial total mass and we leave deeper investigations of the direct cause for future work.} and $Y_\mathrm{c}$ regardless of the mass loss history, i.e., $\delta=\delta(M_\mathrm{zams},\hat{Y_\mathrm{c}})$. Again, we present a fitting function for $\delta(\hat{Y}_\mathrm{c},M_\mathrm{zams})$ in Appendix~\ref{app:formula_fid}.

\subsection{Analytical model for mixing core mass evolution}
In this section, we attempt to construct a self-contained analytical model that can predict the evolution of the mixing core mass given an arbitrary mass loss history. We mainly base our model on Eqs.~(\ref{eq:dmdY_sum}),(\ref{eq:natural_decline_rate}), \& (\ref{eq:dmdy_model}), making use of the relations we discovered in the previous sections. So far we have used the central helium fraction $Y_\mathrm{c}$ as a measure of stellar age to express the core evolution. In practice, the mass loss rate is usually given as a time derivative ($\dot{M}$) and not $dM/dY_\mathrm{c}$. Therefore, we need to convert our model into a time evolution equation. This can be done by multiplying the equations by $\dot{Y}_\mathrm{c}$,
\begin{equation}
	\frac{\mathrm{d} \ln m_\mathrm{mix}}{\mathrm{d} t} = - \frac{\alpha}{1 - \alpha Y_\mathrm{c}} \dot{Y}_\mathrm{c} + \beta(M)\delta(M_\mathrm{zams},\hat{Y}_\mathrm{c}) \frac{\dot{M}}{M}.
	\label{mmix_ev_t}
\end{equation}
The time evolution of the central helium fraction can be expressed as
\begin{equation}
	\dot{Y}_\mathrm{c} = \frac{L}{Q_\mathrm{CNO} m_\mathrm{mix}},\label{eq:dYdt_formula}
\end{equation}
where $Q_\mathrm{CNO}=6.019\times10^{18}$~erg~g$^{-1}$ is the energy released per unit mass by hydrogen fusion via the CNO cycle\footnote{We are interested in massive stars that have convective cores where the CNO cycle is the dominant source of nuclear energy generation.}. 
By integrating the differential equations Eqs.~(\ref{mmix_ev_t}) and (\ref{eq:dYdt_formula}) along with the fitting functions presented in Appendix~\ref{app:formula_fid}: $\alpha(m_\mathrm{mix}), \beta(M), \delta(\hat{Y}_\mathrm{c},M_\mathrm{zams}), L(m_\mathrm{mix}, Y_\mathrm{c})$, we can now compute the mixing core mass evolution given any arbitrary mass loss history $\dot{M}(t)$. The initial mixing core mass is $m_\mathrm{mix} =M_\mathrm{zams} f_\mathrm{mix}(M_\mathrm{zams})$, using the fitting formula for $f_\mathrm{mix}(M_\mathrm{zams})$.

\begin{figure}[ht!]
\includegraphics[width=\linewidth]{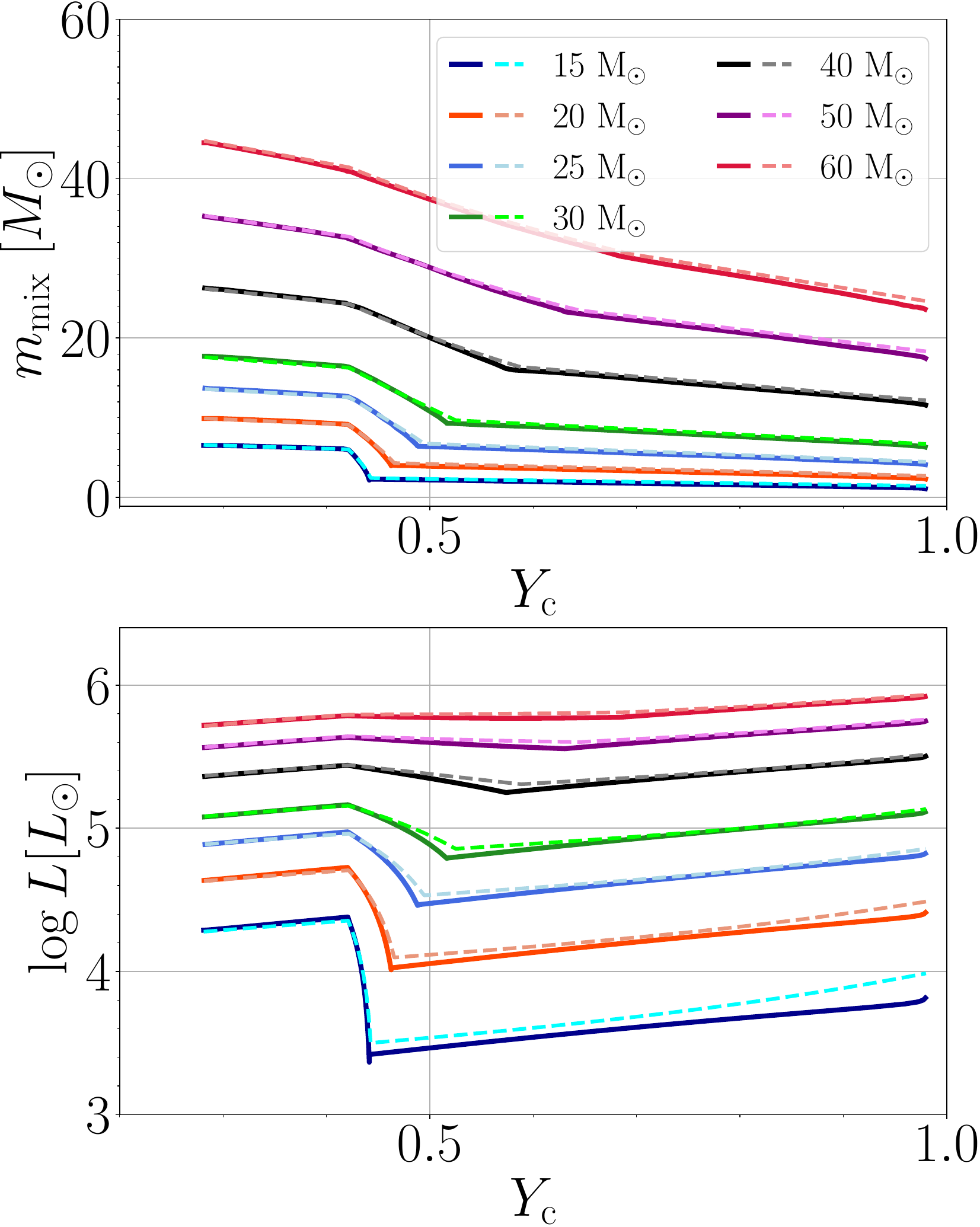}
\caption{Evolution of mixing core mass ($m_\mathrm{mix}$) and luminosity ($L$) against central helium fraction ($Y_\mathrm{c}$) for models with mass loss rate fixed to $\dot{M}=-10^{-5}~\msun~\mathrm{yr}^{-1}$ and mass loss initiated at $\tau_\mathrm{ml}=0.2$.
 The metallicity is set to the solar value.  The solid curves show results of our MESA models whereas the dashed curves show the solutions to the analytical model given by Equation~(\ref{mmix_ev_t}).}
\label{fig:mmix_and_lum_w_different_ini_mass}
\end{figure}

\begin{figure}[ht!]
\includegraphics[width=\linewidth]{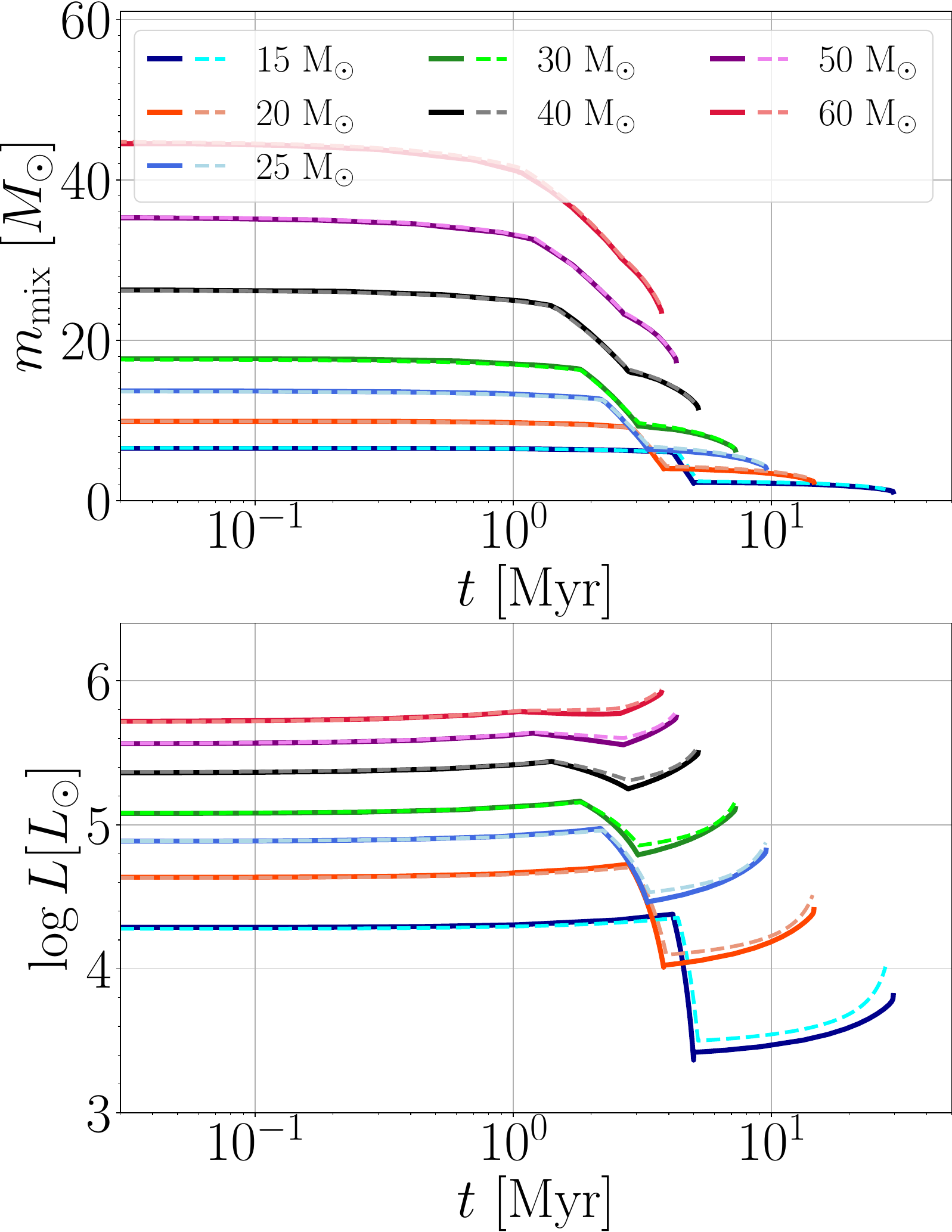}
\caption{Same as Figure~\ref{fig:mmix_and_lum_w_different_ini_mass}, but plotted against age.}
\label{fig:mmix_and_lum_in_time}
\end{figure}

In Figures~\ref{fig:mmix_and_lum_w_different_ini_mass} and \ref{fig:mmix_and_lum_in_time} we show the solutions to these differential equations (dashed curves) along with the actual mixing core mass evolution in our MESA models (solid curves). For the analytical models, we apply the same mass loss history as in the MESA models, in which we switch on a fixed mass loss rate ($\dot{M}=-10^{-5}~\msun~\mathrm{yr}^{-1}$) at the time given by $\tau_\mathrm{ml}=0.2$ and terminate the mass loss once the current total mass reaches the initial mixing core mass ($M=m_\mathrm{mix,zams}$). It is evident that the analytic solution agrees almost perfectly with the MESA results, confirming the validity of our model especially for the mixing core mass evolution (top panels). There are small deviations in the luminosities (bottom panels), particularly for the lower-mass models, despite the mixing core mass mass being in good agreement. This is due to a combination of the poorer quality of fit of our luminosity formula and the natural decline rate parameter $\alpha$ becoming less universal towards the lower mass regions. Apart from the lowest mass model ($M_\mathrm{zams}=15 M_\odot$), the luminosity evolution is sufficiently close to the MESA models so that the \ac{ms} lifetimes are also in perfect agreement, as we can see from the endpoints of each curve in Figure~\ref{fig:mmix_and_lum_in_time}. We find similarly strong agreement between the MESA models and analytical models for all the masses, mass loss rates and mass loss timing combinations we computed.

\section{Discussion} \label{sec:discussion}
\subsection{Parameter dependence}
The universal relations and analytical modelling described above are all based on our fiducial model. However, we find that the same relations apply for other sets of parameter choices (metallicity, mixing parameters, etc.), albeit with slightly different fitting coefficients.
For example, the normalized natural decline rate of the mixing core ($\alpha$) is always universal over all masses within a given set of parameter choices ($Z, \alpha_\mathrm{ov}$). The luminosity is always strongly correlated with $m_\mathrm{mix}$ and $Y_\mathrm{c}$ and nothing else. As expected, the choice of mixing length ($\alpha_\mathrm{mlt}$) does not seem to influence any of our results. Most surprisingly, the ambiguous parameter $\delta$ seems to be consistent with being the same function over all metallicities we simulated.
We summarize the fitting coefficients for the other metallicities considered in Appendix~\ref{app:formula_fid} alongside our fiducial model.

One caveat to be noted is that in all of our models, we assume that the overshoot is a fixed fraction of the local pressure scale height at the convective boundary. There is no strong reason why this should be the case, and in fact some stellar models in the literature \citep[e.g.][]{Pols1998} employ effectively mass-dependent overshoot parameters. In such situations, it is possible that our universal relations do not hold and hence the analytical framework described by Eq.~(\ref{mmix_ev_t}) breaks down. However, we also expect that as long as the amount of overshoot is dependent on only $m_\mathrm{mix}$ and $Y_\mathrm{c}$, the framework should still hold. We leave investigations of more complicated overshooting models for future work.

\subsection{Challenges for implementing Case A mass transfer in rapid binary population synthesis}
So far, using simple stellar models with constant mass loss rates, we showed that our fitting formulae accurately describe the evolution of the convective core mass obtained from MESA. 
This enables us to predict the core mass evolution without carrying out 1D stellar evolution simulations given an arbitrary mass loss history. Utilizing this model, we can greatly improve upon current methods in rapid population synthesis codes in predicting the helium core mass at \ac{tams}, which is critical in a wide variety of contexts, such as X-ray binaries and gravitational-wave sources. Currently for many rapid population synthesis codes based on \citet{Hurleyetal2000}, the core mass at \ac{tams} is predicted based on the total mass at that point, which is equivalent to choosing the endpoint of the grey dashed curve in Figure~\ref{fig:comp_bse}. By comparing this with the other curves in the plot, it is clear that the core mass is underestimated by a significant fraction depending on the exact mass loss history. Our proposed framework can provide a more accurate estimate of the core mass at \ac{tams} for a given mass loss history.

As a by-product, we can also obtain the chemical profile of the star. Figure~\ref{fig:mmix_and_lum_w_different_ini_mass} shows the evolution of the mixing core mass as a function of central helium fraction, but it is equivalent to the internal helium profile of the star at TAMS if we replace $m_\mathrm{mix}$ with mass coordinate $m$ and $Y_\mathrm{c}$ with the local helium fraction $Y(m)$. This may prove useful for other research directions such as in predicting the surface abundances for WN stars \citep{Schootemeijer2018}.

In the context of binary evolution, our formulae only solve half of the problem. Mass transfer from main sequence donors onto companions, so-called Case A mass transfer, is mainly driven by the nuclear evolution of the star. Therefore, unlike other modes of mass transfer that are driven on timescales much shorter than the core evolution (e.g. Case B mass transfer), the mass loss rate is determined by the core evolution itself so both aspects need to be modelled simultaneously. Currently, the mass transfer rates during Case A mass transfer in rapid binary population synthesis codes are computed based on stitching together single MS star models at various effective ages. This is known to be incorrect, as the mass loss history alters the internal chemical profiles, which will be different from single star models.

\begin{figure}[ht!]
\includegraphics[width=\linewidth]{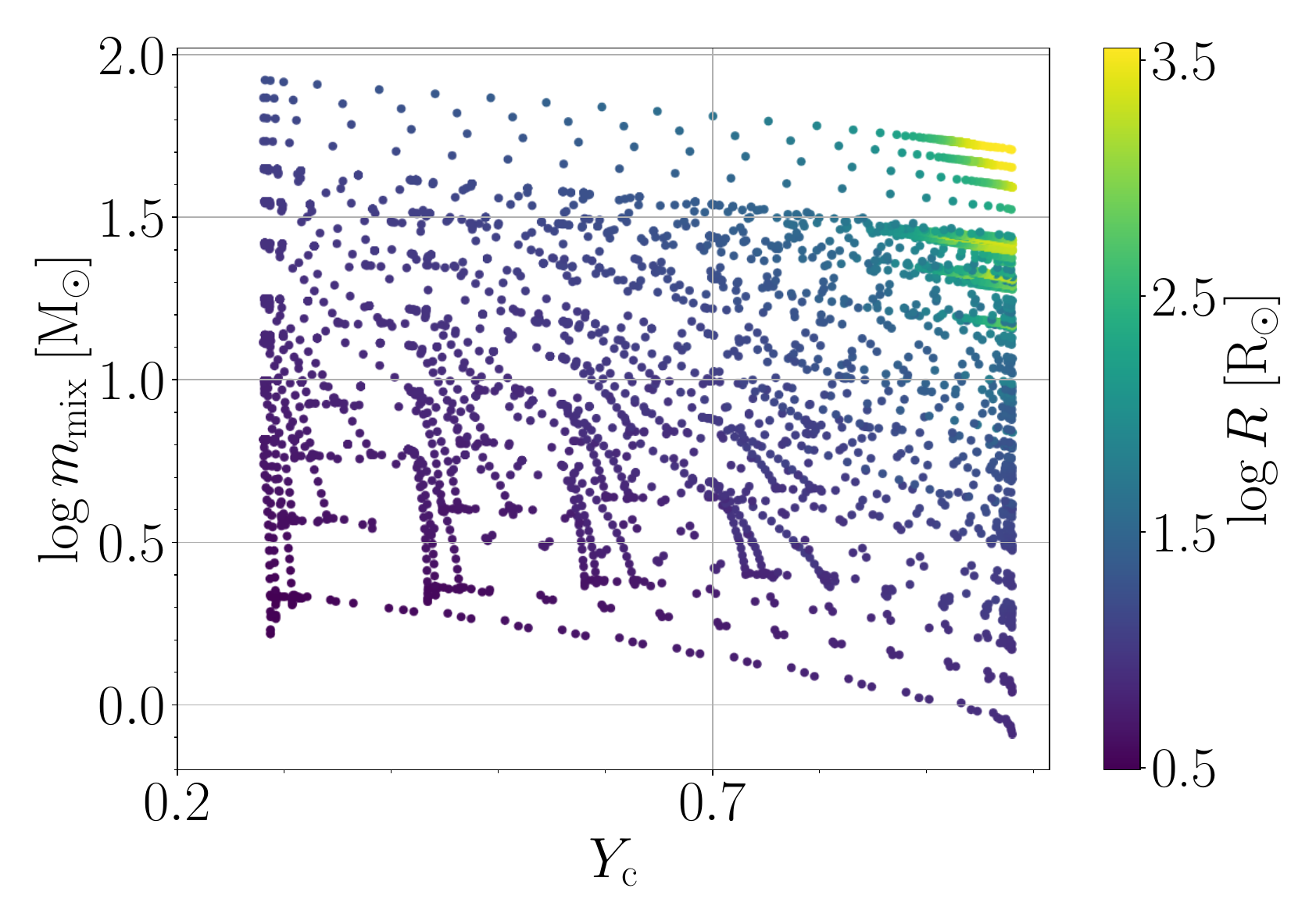}
\caption{Relationship between mixing core mass, central helium fraction and stellar radius (see the caption of Figure \ref{fig:L_mmix}).
\label{fig:R_mmix}}
\end{figure}

The main ingredient required for predicting mass transfer rates is the radius evolution of the star. In Figure~\ref{fig:R_mmix} we show the relations between $m_\mathrm{mix}$, $Y_\mathrm{c}$ and stellar radius $R$. Unlike the luminosity $L$, the radius shows a wide scatter, even for models with similar values of $m_\mathrm{mix}$ and $Y_\mathrm{c}$ (see for example the broad range of radii in the top right corner). This indicates that the radius is sensitive to the details of the mass loss history, or more directly, the full internal chemical profile rather than just the core properties. 
We leave a more thorough investigation of the radial evolution for future work.



\section{Conclusion} \label{sec:conclusion}
Based on simulations with the one-dimensional stellar evolution code MESA, we find universal relationships that describe the convective core mass of MS stars and its evolution with and without mass loss.
Notable relationships are:
\begin{enumerate}
    \item The natural (in the absence of mass loss) fractional decline rate of the mixing core mass is constant regardless of the initial total mass, when mass-independent overshooting schemes are used. The same trend is observed even for stars that had experienced mass loss in the past. 
    \item Stellar luminosity can be expressed by a function of only two parameters: mixing core mass and the central helium fraction.
    \item We define a core \textit{inertia} parameter $\delta$ (Eq.~(\ref{eq:delta})), which is a measure of how the mixing core mass responds to mass loss. This parameter only seems to depend on the central helium fraction and, weakly, on the initial total mass.
\end{enumerate}

The relationships above enable us to formulate an analytical framework to predict the convective core mass evolution given the mass loss history of a star.
The model is based on a differential equation where the decrease of mixing core mass during mass loss is a superposition of the natural decrease of mixing core (characterized by a parameter $\alpha$) and an effect due to the mass loss. The contribution from the latter term can be further separated into two subcomponents; a factor that depends on the total mass ($\beta$) and a factor that depends on the central helium fraction ($\delta$). We present fitting formulae for each of these components ($\alpha, \beta, \delta$) as well as the luminosity. We confirmed that if we solve the differential equation Eq.~(\ref{mmix_ev_t}) using the fitting functions, the evolution of the mixing core mass can be accurately described over a wide range of masses and mass loss histories.


The detailed functional form of the model parameters $\alpha, \beta, \delta$ may vary depending on the choices made for the stellar physics in the 1D stellar evolution calculations (e.g.~metallicity, overshoot). However, we believe our framework for computing the mixing core mass evolution should work for any set of stellar evolution models as long as the parameters are adjusted accordingly. This could be a powerful alternate method to the way \ac{ms} evolution is treated in rapid stellar population synthesis codes \citep{Hurleyetal2000,Agrawal2023,Iorio2023}. It is particularly useful in the massive star regime where the mass loss rates of single stars are very high and stars can lose significant fractions of their mass by the time they reach \ac{tams}. This framework should be useful for modelling mass loss through binary interactions too, although improved models for the radial evolution of mass-losing stars are needed.



\section*{Acknowledgement}
We thank Adam Br\v{c}ek for critical feedback on the manuscript.  This research was funded by the Australian Research Council Centre of Excellence for Gravitational Wave Discovery (OzGrav), through project number CE230100016.

\appendix
\section{Fitting Formulae and Fitting Parameters for the Fiducial Model}
\label{app:formula_fid}

Here, we summarize the fitting formulae required to describe the evolution of the mixing core mass and luminosity.  We provide the fitting parameters for our models with fiducial overshoot parameters ($\alpha_\mathrm{mlt}=2, \alpha_\mathrm{ov}=0.2$).
All masses are in units of solar masses.
We also evaluate the fractional deviation of the fits from the true value (i.e.~the value obtained from MESA) as,
\begin{equation}
    \Delta h \equiv \frac{h_\mathrm{fit} - h_\mathrm{MESA}}{h_\mathrm{MESA}},
\end{equation}
where $h$ is the physical quantity we try to fit, e.g.~$\alpha, f_\mathrm{mix}, \log L$.

\subsection{Natural decline rate}

The value $\alpha$ characterizes the natural decline rate of the mixing core mass in the absence of mass loss and can be expressed as a function of $m_\mathrm{mix}$,
\begin{equation}
    \log (\alpha - a_1) = \max (-2, b_1 m_\mathrm{mix} + c_1).
\end{equation}
where the fitting parameters are summarized in Table~\ref{tab:param_alpha}.
Deviations from the MESA results are within $\lesssim10~\%$ for most of the parameter space except for the lower mass region $m_\mathrm{mix}\lesssim10~\msun$ and timesteps where the star is not in thermal equilibrium.

\begin{table*}[ht!]
	\centering
\begin{tabular}{c|ccc} \hline
metallicity &  $a_1$ & $b_1$ & $c_1$\\ \hline
$Z_\odot$ & $0.45$ & $-0.05878711$ & $-0.84646162$  \\
$\frac{1}{3} Z_\odot$ & \ldots & $-0.06968022$ & $-0.73688164$  \\
$0.1 Z_\odot$ & \ldots & $-0.0557105$ & $-0.86589929$  \\ \hline
\end{tabular}
	\caption{Summary of fitting parameters for $\alpha$. ``\ldots'' means the same value as above.}
 \label{tab:param_alpha}
\end{table*}

\subsection{Initial mixing core mass}

We fit the fraction of the mixing core mass at \ac{zams} $f_\mathrm{mix} \equiv m_\mathrm{mix,zams}/M_\mathrm{zams}$ in Figure~\ref{fig:beta} as a function of the \ac{zams} mass
\begin{equation}
    f_\mathrm{mix} = a_3 + b_3 \exp \left ( - \frac{M_\mathrm{zams}}{c_3} \right ),
\end{equation}
where values of the fitting parameters are listed in Table~\ref{tab:param_fmix}.
We can compute the factor $\beta$ as defined in Eq.~(\ref{eq:beta}) by taking the derivative of $f_\mathrm{mix}$
\begin{align}\label{fmixfit}
    \beta(M_\mathrm{zams}) &= 1 + \frac{\mathrm{d} \ln f_\mathrm{mix}}{\mathrm{d} \ln M_\mathrm{zams}} \nonumber \\
    &= 1 -\frac{b_3 M_\mathrm{zams}}{c_3 f_\mathrm{mix}} \exp{\left ( - \frac{M_\mathrm{zams}}{c_3} \right )}.
\end{align}
Again the fitting parameters are summarized in Table~\ref{tab:param_fmix}.
The maximum deviation of the fit from the true value is $\Delta f_\mathrm{mix}\leq0.01~\%$.

\begin{table*}[ht!]
	\centering
\begin{tabular}{c|ccc} \hline
metallicity & $a_3$ & $b_3$ & $c_3$\\ \hline
$Z_\odot$ & $0.86605495$ & $-0.64960375$ & $35.57019104$ \\
$\frac{1}{3} Z_\odot$ & $0.86269445$ & $-0.62623353$ & $35.74630996$  \\
$0.1 Z_\odot$ & $0.86914766$ & $-0.60815098$ & $37.20654856$  \\ \hline
\end{tabular}
	\caption{Summary of fitting parameters for $f_\mathrm{mix}$.}
 \label{tab:param_fmix}
\end{table*}

\subsection{Luminosity}

The stellar luminosity (in units of solar luminosity $L_\odot$) can be expressed as a function of only $m_\mathrm{mix}$ and $Y_\mathrm{c}$
\begin{align}
        \log L = &a_4 \log m_\mathrm{mix} + b_4 Y_\mathrm{c} + c_4 (\log m_\mathrm{mix}) Y_\mathrm{c} + \nonumber \\
        &d_4 (\log m_\mathrm{mix})^2 + e_4 Y_\mathrm{c}^2 + f_4 (\log m_\mathrm{mix})^3 + g_4 Y_\mathrm{c}^3 + \nonumber \\
        &h_4 (\log m_\mathrm{mix})^2 Y_\mathrm{c} + i_4 (\log m_\mathrm{mix}) Y_\mathrm{c}^2 + j_4,
\end{align}
where the fitting parameters are summarized in Table~\ref{tab:param_ltot}.
The fitting functions are accurate to within $\Delta L\lesssim 2$~\% over all the MESA models we simulated.

\begin{sidewaystable}[ht]
\centering
\begin{tabular}{c|cccccccccc} \hline
metallicity & $a_4$ & $b_4$ & $c_4$ & $d_4$ & $e_4$ & $f_4$ & $g_4$ & $h_4$ & $i_4$ & $j_4$\\ \hline
$Z_\odot$ & $3.27883249$ & $1.79370338$ & $-0.71413866$ & $-0.77019351$ & $-0.3898752$ & $0.07499563$ & $0.5920458$ & $0.33846556$ & $-0.49649838$ & $1.71263853$  \\
$\frac{1}{3} Z_\odot$ & $3.35622529$ & $1.96904931$ & $-0.88894808$ & $-0.81112488$ & $-0.47925922$ & $0.09056925$ & $0.53094768$ & $0.33971972$ & $-0.35581284$ & $1.65390003$ \\
$0.1 Z_\odot$ & $3.2555795$ & $1.84666823$ & $-0.79986388$ & $-0.75728099$ & $-0.38831172$ & $0.08223542$ & $0.49543834$ & $0.31314176$ & $-0.36705796$ & $1.72200581$ \\ \hline
\end{tabular}
	\caption{Summary of fitting parameters for $L$.}
 \label{tab:param_ltot}
\end{sidewaystable}

\subsection{Mixing core inertia parameter}
The value of the mixing core \textit{inertia} parameter $\delta(M_\mathrm{zams},\hat{Y}_\mathrm{c})$ can be fit by
\begin{align}
    \delta(M_\mathrm{zams},\hat{Y}_\mathrm{c}) &= \min (10^{-\hat{Y}_\mathrm{c} + g(M_\mathrm{zams})}, 1), \\
    g(M_\mathrm{zams}) &= a_2  M_\mathrm{zams} + b_2,
\end{align}
where $a_2 = -0.0044$, $b_2 = 0.27$.
The typical deviation is $\Delta\delta\lesssim10$~\%.


\bibliography{sample631}{}
\bibliographystyle{aasjournal}



\end{document}